\documentclass[conference]{IEEEtran}
\IEEEoverridecommandlockouts
\usepackage[noadjust]{cite}

\ifCLASSINFOpdf
  \usepackage[pdftex]{graphicx}
\else
\fi
\usepackage[cmex10]{amsmath}
\usepackage[utf8]{inputenc}
\usepackage{pstricks}
\usepackage{array}
\usepackage{amsmath}
\usepackage{pifont}
\usepackage{dsfont}
\usepackage{multirow}
\usepackage{amssymb}
\usepackage{wasysym}

\def\be{\begin{equation}}
\def\ee{\end{equation}}

\newcommand\figref{Fig.~\ref}
\begin{document}
%
\title{Classical and quantum spreading of a charge pulse}

\author{\IEEEauthorblockN{B. Gaury, J. Weston, C. Groth, and X. Waintal}
\IEEEauthorblockA{Univ. Grenoble Alpes, INAC-SPSMS, F-38000 Grenoble, France}
\IEEEauthorblockA{CEA, INAC-SPSMS, F-38000 Grenoble, France}
\IEEEauthorblockA{e-mail: xavier.waintal@cea.fr}
}

\IEEEpubid{\makebox[\columnwidth]{\hfill 978-1-4799-5433-9/14/\$31.00~\copyright~2014~IEEE} \hspace{\columnsep} \makebox[\columnwidth]{ }}%


%


\maketitle

\begin{abstract}
With the  technical progress of radio-frequency setups, high frequency quantum
transport experiments have moved from theory to the lab. So far the standard
theoretical approach used to treat such problems numerically---known as Keldysh
or NEGF (Non Equilibrium Green's Functions) formalism---has not been very
successful mainly because of a prohibitive computational cost. We propose a
reformulation of the non-equilibrium Green's function technique in terms of the
electronic wave functions of the system in an energy--time representation. The
numerical algorithm we obtain scales now linearly with the simulated time and
the volume of the system, and makes simulation of systems with $10^5-10^6$
atoms/sites feasible. We illustrate our method with the propagation and
spreading of a charge pulse in the quantum Hall regime. We identify a classical
and a quantum regime for the spreading, depending on the number of particles
contained in the pulse.  This numerical experiment is the condensed matter
analogue to the spreading of a Gaussian wavepacket discussed in quantum mechanics
textbooks.
 \end{abstract}


%
\IEEEpeerreviewmaketitle

\section{Introduction}
Finite-frequency quantum transport and particle physics are somehow similar;
accessing higher frequencies unlocks new physics as the probing frequency
crosses characteristic frequencies of the quantum systems. The first of these is
the temperature---$\hbar\omega = k_B T$ translates into 20~GHz
$\leftrightarrow$ 1~K---and the recent technical progress in assembling
radio-frequency lines ($\sim 10~\mathrm{GHz}$) in dilution fridges ($\sim
10~\mathrm{mK}$) have made the domain of time-resolved quantum transport
accessible in the lab. This opens the door to the manipulation of very few or
even single electrons, a key step towards the development of single electron
sources, and in a broader context towards quantum computing. Two different
techniques to realize such a source have given promising results. In the first
one, an AC signal is applied to a quantum dot which then releases one particle
into the connected Fermi sea~\cite{Single_e_source}. This procedure allows a
fine control of the energy of the particles, but not of their releasing time.  A
second route taken in~\cite{Glattli2013} consists of applying a voltage pulse to
an Ohmic contact.  This technique offers better control of the time-dependence
of the source but creates excitations in a wide range of energies. In a
single-mode system in the linear regime, such a voltage pulse $V(t)$ induces a
current $I(t)=(e^2/h)V(t)$, injecting $\bar n=\int eV(t)/h$ particles.  These
sources have been, up to now, mainly used in the reproduction of known quantum
optics experiments~\cite{Bocquillon2013}. Electrons, however, are not photons and
we anticipate many new effects when using the former. In particular, the Fermi
sea, always present with fermions but absent in bosonic systems, plays a crucial
role.

In this manuscript, we first review our wave-function approach
(section~\ref{theory}) and then focus on a very simple problem: the spreading of
a charge pulse in a one-dimensional system (section~\ref{spreading}). For
practical numerical calculations, we simulate the latter using the edge states
of a two-dimensional gas in the quantum Hall regime. Therefore the numerics are
actually done on a 2D system. In this context, the charge pulses are closely
related to the so-called edge magnetoplasmons which have been studied for a long
time~\cite{glazman1994, Kumada2011, Petkovic2013}. 

\section{From NEGF to a wave function approach}
\label{theory}
We start with the main results of the Keldysh formalism and completely
reformulate it by introducing a time-dependent wave function. We refer
to~\cite{Cookbook_Oleksii,Rammer_review, Rammer_book} for more details on the
NEGF formalism, and to~\cite{Twave_formalism} and references therein for a
derivation of the wave function approach and the numerical implementation.

\begin{figure}[h]
    \center
    \includegraphics[width=0.48\textwidth]{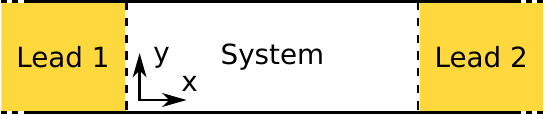}
    \caption{\label{system} Schematic of an open system connected to two leads
    that are kept at equilibrium with
    temperature $T_{1/2}$ and chemical potential $\mu_{1/2}$.}
\end{figure}

We model an open system with  the tight-binding Hamiltonian
\begin{equation}
    \mathrm{\hat{\textbf{H}}}(t) =
    \sum_{i,j} \mathrm{\textbf{H}}_{ij}(t) c^{\dagger}_{i}c_{j},
    \label{ham}
\end{equation}
where $c^{\dagger}_i$ ($c_j$) are the Fermionic creation (annihilation) operators of a
one-particle state on site $i$. The system consists of a central region
connected to two leads as depicted in \figref{system}. We chose a quasi
one-dimensional system for illustrative purposes, but the formalism is
completely general and allows the treatment of any multi-terminal device.
The basic objects that we consider are the
Retarded ($G^R$) and Lesser ($G^<$) Green's functions defined in the central
region $\bar 0$. Integrating out the degrees of freedom of the leads, one
obtains effective equations of motion for $G^R$ and
$G^<$~\cite{Rammer_review, Rammer_book},
\begin{align}
\label{eq:EOM}
& i \partial_{t} G^{R}(t,t') =
 \mathrm{\bold{H}}_{\bar{0}\bar{0}}(t)G^{R}(t,t')
 + \int du\ \Sigma^{R}(t,u)G^{R}(u,t')\\
&G^< (t,t') = \int du \int dv\ G^{R}(t, u) \Sigma^{<}(u,v) [G^{R}(t',v)]^{\dagger}
\label{eq:EOM<}
\end{align}
with the initial condition for $G^R$ given by $G^R(t',t') = -i$. The
self-energies are spatial boundary conditions that take into account the effect
of the leads
\be
\Sigma^k(t,t') = \sum_{\bar m=1}^{\bar M} \Sigma^k_{\bar m}(t,t'),\ \ \ \ k=R, <,
\ee
where $\Sigma_{\bar{m}}^k$ is the self-energy of lead $\bar{m}$. The
calculation of observables, such as the particle current or the local
electronic density, amounts to first solving the integro-differential
equation Eq.~(\ref{eq:EOM}), followed by the double integral of
Eq.~(\ref{eq:EOM<}). Physical quantities are then written in terms of the Lesser
Green's function. We note $N$ the number of sites inside the system and $S$ the
number of sites at the system--lead interfaces. We also note $t$ the maximum
time of a simulation and $h_t$ the typical discretization time step. A naive
integration of Eq.~(\ref{eq:EOM}) and Eq.~(\ref{eq:EOM<}) scales as
$(t/h_t)^2S^2N$. Such a scaling makes the NEGF formalism very demanding from a
computational viewpoint. In addition, the integration of Eq.~(\ref{eq:EOM<}) is
problematic as the integral converges slowly and large computing times are used
simply to recover the equilibrium properties of the system. 

The wave function approach allows one to solve these difficulties.  One first
obtains the stationary wave function $\Psi_{\alpha E}^{st}$ of the system (using
for instance the Kwant package developed by some of us\cite{kwant_preparation})
and simply solves the Schrödinger equation,
\be
\label{wbl2}
   i \partial_{t} \Psi_{\alpha E}(t) = \mathrm{\textbf{H}} (t) \Psi_{\alpha E}(t)
\ee
with the initial condition 
\be \label{wbl3}
\Psi_{\alpha E}(t<0)=\Psi_{\alpha E}^{st}e^{-iEt}
\ee 
(where we have supposed for convenience that the time-dependent perturbation is
only present for $t>0$). The Lesser Green's function, hence the physical
observables (density, current, ...) are then simply expressed in terms of these
wave functions,
\be
    G^<(t,t') =\sum_\alpha \int \frac{dE}{2\pi}\ if_\alpha(E)  \Psi_{\alpha
    E}(t) \Psi_{\alpha E}(t')^\dagger,
    \label{eq:psi-less}
\ee
where $f_{\alpha}(E)$ is the Fermi function of lead $\alpha$. In practice, one
considers the deviation to the stationary solution and introduces, 
\be
\bar \Psi_{\alpha E}(t) = {\Psi}_{\alpha E}(t) - e^{-iEt} \Psi_{\alpha E}^{st},
\ee
${\bar \Psi}_{\alpha E}(t)$ satisfies,
\be
\label{wbl}
   i \partial_{t} \bar\Psi_{\alpha E}(t) = W (t) \bar\Psi_{\alpha E}(t) +
   S(t),
\ee
with $\bar \Psi_{\alpha E}(t<0)=0$ and a ``source'' term localized where the
time-dependent perturbation takes place, 
\be
S(t) = [\mathrm{\textbf{H}}(t) - \mathrm{\textbf{H}}(t=0)] e^{-iEt}\Psi_{\alpha
E}^{st}.
\ee
Once the wave function starts spreading in the lead, it never comes back, the
leads are invariant by translation and therefore full transmitting. The leads
are therefore taken into account (exactly) by using absorbing boundary
conditions. Eq.~(\ref{wbl}) can be solved very efficiently in parallel for the
different energies and modes.

\begin{figure}[h]
    \includegraphics[width=0.48\textwidth]{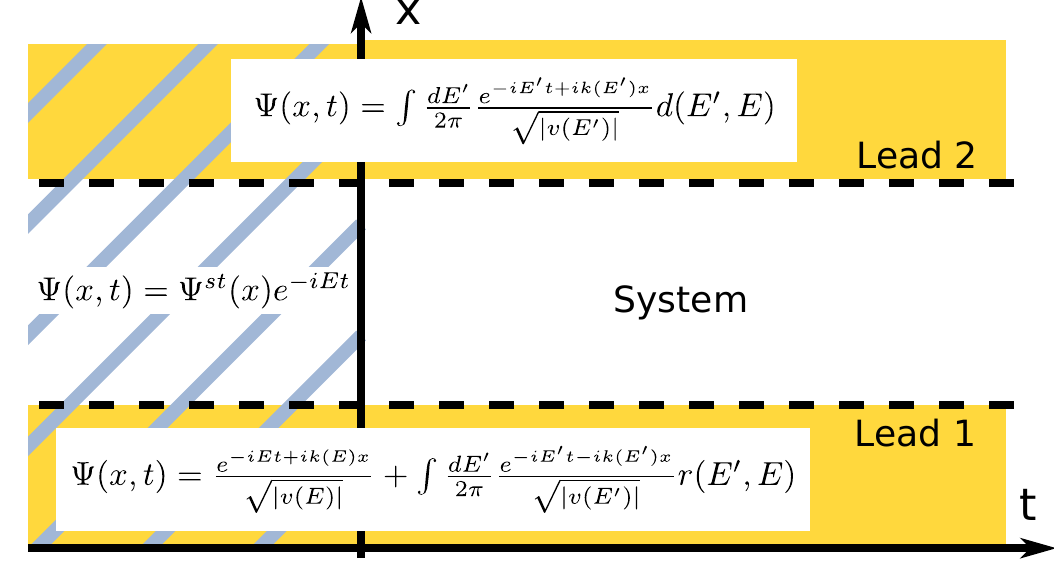}
    \caption{Boundary conditions for the scattering matrix theory (yellow) and
    the partition-free approach (blue lines). }
    \label{BC}
\end{figure}

\figref{BC} illustrates the various approaches taken to describe time-dependent
transport. One can consider Green's functions or wave functions, but one can
also consider two, apparently different, boundary conditions. In the first
[described above by Eq.~(\ref{wbl2}) and (\ref{wbl3}) ] the boundary condition
is given for all $x$ and $t=0$ (this is known in the literature as the
partition-free approach). In the second, the scattering approach, one imposes
the form of the wave function in the leads (fixed $x$) at all times $t$. Both
approaches are in fact identical. 

\section{Spreading of a charge pulse in the quantum Hall regime}
\label{spreading}
\begin{figure}[h]
    \includegraphics[width=0.48\textwidth]{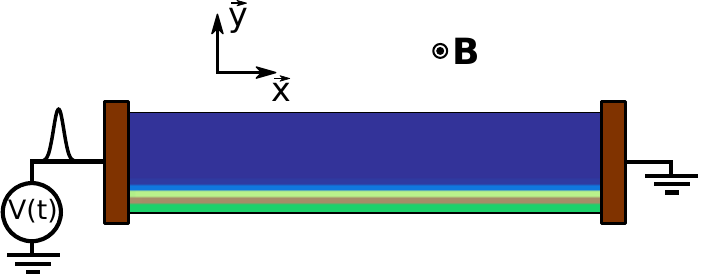}
    \caption{Color map of $\partial \rho(x,y)/ \partial V$ of the two-dimensional
    electron gas showing the position of the edge state at the Fermi energy.}
    \label{2DEG}
\end{figure}

We now apply the formalism that has been introduced in the previous section to
the spreading of a charge pulse in the quantum Hall regime.  We consider a
two-dimensional electron gas (2DEG) under high magnetic field connected to two
Ohmic contacts as depicted in \figref{2DEG}.
We work in a regime where the transport properties are fully determined by the
lowest Landau levels (LLL). We send voltage pulses via the left contact while the
right one is grounded.  \figref{2DEG} is not a simple schematic of the system,
but shows the electronic charge distribution $\partial \rho(x,y)/ \partial V$
appearing in the 2DEG upon applying a DC bias voltage $V$ at the left contact.
The Hamiltonian for the system reads
\be
\label{eq:H}
      \mathrm{\hat{\textbf{H}}} =
      \frac{(\vec P -e\vec A)^2}{2m^*} +V(\vec r,t),
\ee
where $\vec P=-i\hbar\vec\nabla$, and $\vec A = -B y \vec x$ is the vector
potential in the Landau gauge.  $B$ is the magnetic field and $m^*$ is the
electron effective mass. $V(\vec{r}, t)$ contains the voltage pulse applied to
the left Ohmic contact and the confining potential due to the mesa boundary.
Equation~(\ref{eq:H}) is discretized on a lattice following standard
practice~\cite{Knit} with parameters corresponding to a GaAs/AlGaAs
heterostructure. We use an electronic density $n_s=10^{11}~\mathrm{cm^{-2}}$
which gives a Fermi energy $E_F=3.47~\mathrm{meV}$ (or a Fermi wave length
$\lambda_F=79~\mathrm{nm}$). We take a magnetic field $B=1.8~\mathrm{T}$ that
corresponds to a magnetic length $l_B=19~\mathrm{nm}$, and the width of the
system is $150~\mathrm{nm}$.

\begin{figure}[h]
    \includegraphics[width=0.48\textwidth]{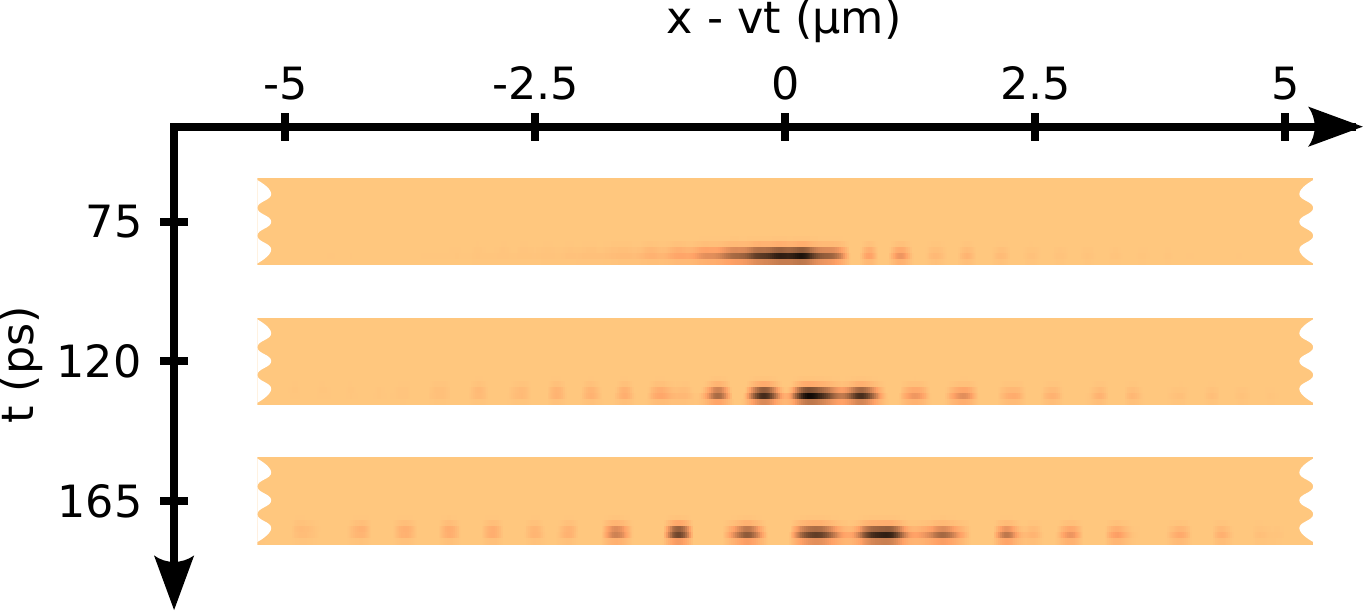}
    \caption{Charge density color map of the spreading of a charge pulse generated by a
    Lorentzian voltage pulse, $V(t)=V_p/(1 + (t/\tau_p)^2)$, with amplitude
    $V_p=0.5mV$ and duration $\tau_p=5ps$.}
\label{colorplots}
\end{figure}

\begin{figure}[b!]
    \includegraphics[width=0.48\textwidth]{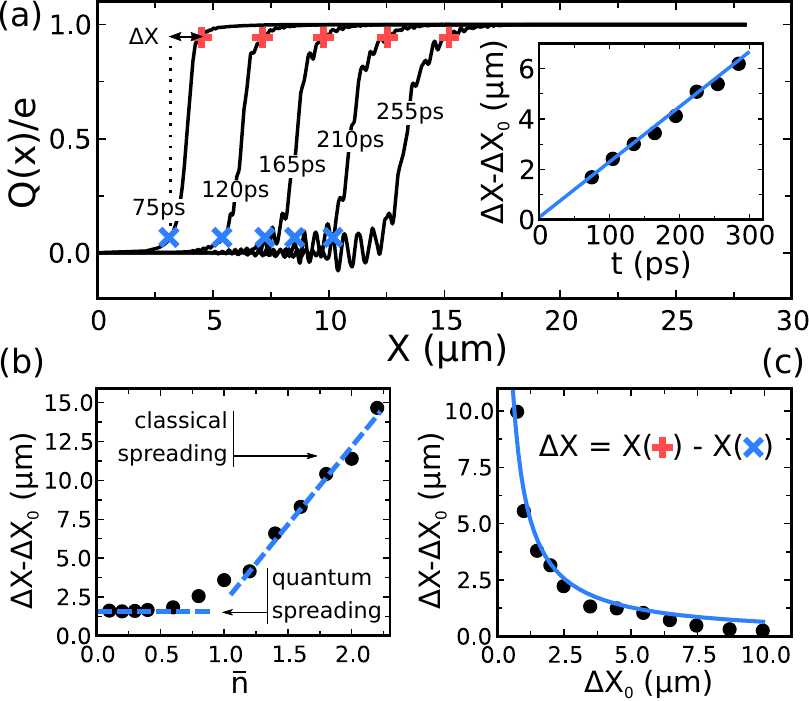}
    \caption{ (a) Number of particles as a function of space (integrated along
    the y-direction). Symbols
    correspond to $5\%$ (blue cross) and $95\%$ (red pluses) of the particles sent.
    Inset: spreading of the charge pulse as a function of time. The full
    line is a linear fit $\Delta X - \Delta X_0=at$. (b) Spreading
    of the charge pulse as a function of the number of particles sent $\bar n$.  The dots
    correspond to numerical data and the dashed
    blue lines guide the eye to distinguish between the quantum and the
    classical regime. (c) Spreading of the charge pulse as a function of its
    initial spatial extension. The dots are numerical data and the
    continuous line correspond to the fit $\Delta X - \Delta X_0=a/\Delta X_0$.
    Parameters for the Lorentzian voltage pulse: (a) $\tau_p=5~\mathrm{ps}$, $\bar n=1$, (b)
    $\tau_p=5~\mathrm{ps}$, (c) $\bar n=1$, with $\bar n = (e / h) V_p \tau_p /4$. (b) and (c) are
    calculated at $t=200~\mathrm{ps}$.}
\label{deltaX}
\end{figure}

\figref{colorplots} shows the propagation of a charge pulse generated by a
Lorentzian voltage pulse $V(t)=V_p/(1 + (t/\tau_p)^2)$, with amplitude
$V_p=0.5~\mathrm{mV}$ and duration $\tau_p=5~\mathrm{ps}$, applied to the left
contact. We represented the deviation of the electronic charge from equilibrium
in the center of mass of the pulse at three different times. The corresponding
charge integrated along the y-direction is plotted in \figref{deltaX}a. One observes (i)
a ballistic propagation at the Fermi group velocity, (ii) a global spreading of
the charge pulse and (iii) oscillations of charge density inside its envelope.  A
similar feature was already shown in~\cite{Twave_formalism} for
the propagation of different shapes of voltage pulses (Lorentzian and Gaussian)
in a one-dimensional chain.
We study the propagation of the pulse in the 2DEG within a Landauer-Büttiker
approach using the concept of one-dimensional edge states~\cite{Buttiker1988}.
The system is invariant by translation in the x-direction, hence in absence of
voltage pulse the LLL are eigenstates of the Hamiltonian Eq.~(\ref{eq:H}) with the plane waves
\be
\label{hello}
    \psi_k(x,y, t) = e^{-(y - kl_B^2)^2 / 4l_B^2}\ e^{ik x}.
\ee
Following the results
obtained for a one-dimensional chain~\cite{Twave_formalism}, we see that
in presence of the voltage pulse, $\psi_k$ becomes $\psi=Y\psi_k$.
In the case of a Lorentzian pulse an explicit expression can be
obtained for the modulation $Y$ in the coordinate of the center of mass of the
charge pulse $X=x-vt$,
\begin{align}
Y(X,t) = 1 - v\tau_P & \sqrt{\frac{2m^*\pi}{it}} \exp\left( \frac{m^* (iX -
v\tau_P)^2}{2it}\right) \nonumber\\ &\times \Bigg[ 1 + {\rm Erf}
\left(\frac{iX-v\tau_P}{2\sqrt{it/(2m^*)}}\right) \Bigg]
\label{eq:Ylor}
\end{align}
with $v=\partial E/\partial k$ and  the usual definition of the error function ${\rm Erf}(x) =
(2/\sqrt{\pi})\int_0^x e^{-u^2}du$. It can be shown from Eq.~(\ref{eq:Ylor})
that the oscillations observed in \figref{colorplots} spread
diffusively~\cite{Twave_formalism}  $\propto\sqrt{t}$ according to the expected
behavior of a wave packet in quantum mechanics. However, the width $\Delta
X(t)$ of the envelope of the charge density spreads {\it linearly} in time.
$\Delta X(t)$ can be obtained analytically from the exponential decay of $|Y|^2$
with $X$ or numerically by looking at the envelope of the electronic density
$\rho(x,y,t)$. 
In practice, we calculate $Q(x,t)=\int dy\int_0^x d\bar x
\rho(\bar x,y,t)$ and define $\Delta X$ as the difference between the blue and
red crosses in \figref{deltaX}a.  We identify two contributions to the spreading
as can be seen in \figref{deltaX}b. First we expand the exponential argument in
Eq.~(\ref{eq:Ylor}) and find that the spatial extension of the envelope of the
charge pulse $\Delta X|_{qu}$ is typically given by
\be
    \Delta X\Big|_{qu} = \frac{t}{m^* \Delta X_0},
    \label{deltaqu}
\ee
with $\Delta X_0=v\tau_p$ the initial spatial extension of the pulse.
Fig.~\ref{deltaX}b shows that Eq.~(\ref{deltaqu}) is valid only in the quantum
regime that is bounded by $\bar n \approx 1$. We shall also consider a
``hydrodynamic'' aspect of the spreading. This second contribution arises when
one considers how the various states $\psi_k$ are filled (with Fermi statistics).
Upon varying the potential on the left Ohmic contact between $0$ and $V_p$, one
injects particles with different energies and hence different velocities into
the system.  To first order in $V_p$, we find that the difference of speed
between the fastest and the slowest particles is given by $V_p/(vm^*)$. We
recast the amplitude of the voltage pulse in terms of the number of particles it
contains $\bar n \sim V_p \tau_p$. This yields a ``classical'' component of the
spreading of the charge pulse,
\be
    \Delta X\Big|_{cl} = \frac{\bar n t}{m^*\Delta X_0}.
    \label{deltacl}
\ee
The second part of Fig.~\ref{deltaX}b ($\bar n > 1$) confirms the scaling of
Eq.~(\ref{deltacl}) with the number of particles injected by the voltage pulse.
Overall \figref{deltaX}c confirms the scaling in $1/\Delta X_0$ of
Eqs.~(\ref{deltaqu}) and~(\ref{deltacl}).

\section{Conclusion}
Fast quantum electronics is still an emerging field, both experimentally
and, to some extent, theoretically. Our formalism paves the way for the simulation of
systems of large size, enabling us to target the physical scales relevant
to actual mesoscopic devices. We have shown
that the transport properties of a voltage pulse applied to an Ohmic
contact are closely related to its quantum nature, that is already expected to
exhibit intriguing results~\cite{fabryperot}.


\section*{Acknowledgment}
This work is funded by the ERC consolidator grant MesoQMC.

\newpage



%

\bibliographystyle{IEEEtran}
\bibliography{references}

\end{document}